 \documentclass[final,5p,times,twocolumn]{elsarticle}

\usepackage{amssymb}
\usepackage{lipsum}

\usepackage{graphicx}
\usepackage{caption}
\usepackage{subcaption}
\usepackage{lipsum}

\journal{IEEE Transactions}

\begin{document}

\begin{frontmatter}



\title{Machine Learning for Dynamic Management Zone in Smart Farming}


\author[first]{Chamil Kulatunga}
\affiliation[first]{organization={University College Dublin},
            city={Dublin},
            country={Ireland}}
\author[second]{Sahraoui Dhelim}
\affiliation[second]{organization={University College Dublin},
            city={Dublin},
            country={Ireland}}
\author[third]{Tahar Kechadi}
\affiliation[third]{organization={University College Dublin},
            city={Dublin},
            country={Ireland}}

\begin{abstract}
Digital agriculture is growing in popularity among professionals and brings together new opportunities along with pervasive use  of modern  data-driven technologies. Digital agriculture approaches can  be used to replace all traditional  agricultural system at very reasonable  costs.  It is  very effective  in optimising large-scale  management of resources, while traditional techniques cannot even  tackle the problem. In this paper, we proposed a dynamic management zone delineation approach based on Machine Learning clustering algorithms using crop yield data, elevation and soil texture maps and available NDVI data. Our proposed dynamic management zone delineation approach is useful for analysing the spatial variation of yield zones. Delineation of yield regions based on historical yield data augmented with topography and soil physical properties helps farmers to economically and sustainably deploy site-specific management practices identifying persistent issues in a field. The use of frequency maps is capable of capturing dynamically changing incidental issues within a growing season. The proposed zone management approach can help farmers/agronomists to apply variable-rate N fertilisation more effectively by analysing yield potential and stability zones with satellite-based NDVI monitoring.



\end{abstract}




\end{frontmatter}




\section{Introduction}
\label{introduction}

Agriculture 4.0 is using many modern research and technologies in different aspects of agriculture including genomics, nanotechnology, synthetic proteins, Internet of Things, automation and machine learning \cite{laurens}. As an important pillar in this space, data-driven agriculture has gain a momentum in last twenty years as a retrofitting mechanism for the available technologies to feed 9 billion population in 2050. It has become more realistic than ever due to wider use of sensors, cloud computing and their integration with cyber-physical-social farming systems to use big data for intuition, intelligence and insights. However, data-driven agriculture is challenging for small actors but important for global sustainability compared to others industries such as healthcare, fin-tech and manufacturing. Those challenges come with small profit margins, climate change activities, ever decreeing land and labour. But certainly data-driven systems in agriculture sheds some light on sustainable intensification in agriculture to reduce environmental footprint and to maximise economic returns \cite{evagelos}.  

Arable farming contributes considerably to the world food production as cereal a main staple food and also as a sustainable crop in different climatic regions in the world. Several new technologies for better data collections are becoming available in crop framing such as soil scans, remote and proximal crop growth sensing, yield quality and quantity monitoring, granular weather monitoring, etc. Arable fields naturally have contiguity of those data of within-field variations for regionalization based on homogeneous sub-fields \cite{lark1}. Farmers and agronomists look into site-specific management practices of large fields by capturing time and spatial variability. Due to economic and logistic reasons, soil sampling are not frequent enough to understand its impact on annual yield. For example P, K, Mg are tested once for three years. However, altitude, soil texture data are not changed or changed slowly. Based on our data management experience in UK farms, yield maps are being collected by many farmers in the last two decades. Most of the analyses have been focused on spatial variability of individual maps. Due to lack of consecutive number of yield maps and crop rotation complexities, both spatio-temporal analysis has been limited so far \cite{gerald}. Therefore, many farmers, agronomists and scientists are interested in looking at the relations of those data layers, deriving compound new data layers and accordingly make site-specific decisions in seeding, fertilization, sub-soiling etc \cite{mariano}.

Ping et. al. \cite{ping} and Luck et. at \cite{luck} focused on cleaning data and analysing a single yield map to identify different zones using uni-variate statistical techniques . Identifying yield potential areas with precise zone boundaries helped to understand issues with current growing season. However, by stacking a set of yield maps in the past years to generate a yield frequency map helped identifying more persistent issues within a field such as soil compaction, erosion, water logging etc \cite{diker}, \cite{lark2}. With farmers' local knowledge and agronomists' expertise, they make some useful decisions based on those homogeneous zones. Later studies \cite{javier} combined those maps with other auxiliary data layers using multi-variate techniques. Leroux et al. \cite{leroux} developed a stable temporal and spatial yield potential/response maps. The authors have used co-occurrence matrix and image textural analyses to assess temporal stability using limited data from two fields in France and UK which grown wheat and canola in 2003-2015 period. It uses seeded region growing algorithm to zone delineation and multivariate Euclidean distance to generated variance map from the yearly yield maps. A variance reduction-based approach is used to evaluate the relevance of zoning. Two Haralick indices (sum of squares and sum of average) are used to characterise the zones. Results are discussed based on rainfall and soil texture characteristics. 

Some multivariate approaches for delineation of management zones have been proposed mostly based on linear statistical techniques. Gerald Blasch et al. \cite{blach} have devised a pattern recognition-based approach called Multi-temporal Yield Pattern Analysis (MYPA) using Principal Component Analysis (PCA). Regions are identified as productive (by the mean of yield points) and stable (by the variance) by stacking 10+ years of yield maps. PCA is used to identify patterns of similarities and differences of normalised multi-year yields. Finally, the process delineates productivity-stability units, which are easy to manage and coherent, based on fuzzy k-means clustering. Zonal Opportunity index is used to find the optimal number of zones. Yield patterns have also been integrated with soil texture, rainfall patterns and decisions from some management practices. 

Bruno Basso et al. \cite{basso} at the Michigan State University (MSU) derive yield stability classes that responds differently to N using remote sensing data. Their work investigates the loss of N in agriculture fields based on high and low stable yield zones. These stability zones are delineated based on NDVI from non-commercial satellite data.  Both the impact of this loss into the environment and (financially) to the farmers are analysed. Eight years of NDVI is used to identify yield stability classes based on year-to-year variability (from mean NDVI of a field). Sub-field yield is estimated based on deconvolving county-level yield data and NDVI demarcated regions. They used minimum of 3 years yield maps. 

G. Buttafuoco et al. \cite{buttafuoco} have used factorial co-kriging as the first regionalised factor to delineate homogeneous management zones as three iso-frequency classes. Authors use polygon kriging to finalise the regions as stable (by means and variances) zones also by integrating with other factors. This process takes into account the spatial autocorrelation of in-field variations of multiple variables (P, SOM, N, field capacity, wilting point, clay and sand). This work uses co-kriging as a robust multivariate geostatistical technique, which uses a weighting function with geo-coordinates. 

Availability of both remote and proximal sensing technologies to measure current crop growth using different Vegetation Indexes (VI) has increased \cite{kayad}, \cite{geogi}. It is also accepted that fertilizer applications like Nitrogen (N) in yield unstable areas should be conducted based on current crop growth \cite{basso}. Therefore differentiation of recent crop growth based on historical or within season performance is useful. In this paper we use yield FMs and VI maps to identify persistently high/low yield zones and incidentally high/low growths. This will help apply N fertilizer optimally within a field using an variable-rate applicator. The approach uses regression based spatially varying coefficients approach to identify regions based on most influential factors locally.

Several statistical methodologies have been used in the literature \cite{gelfand}, \cite{margaret} to analyse spatial variability in agricultural fields. Among them geo-statistical methods using spatial heterogeneity and spatial correlations \cite{long}, \cite{koutsos}, \cite{evans} have shown very good potentials. Moran local index, Geographically Weighted Regression, Eigenvector Spatial Filtering are the handful of techniques in geo-spatial analytics. Therefore, we use Moran's LISA clustering for yield region identification and Geographically Weighted Regression (GWR) for identifying influential factor within a region.

\section{Materials and Methods}

\subsection{Study Area and Data Formation}
Data used in this study is based on 3 cereal growing farms from the main Winter Wheat growing region of eastern and southern England. All three farms grew winter or spring varieties of wheat, oil seed rape, barley or beans in rotations on rain-fed (rainfall concentrated on winter months) lands. Three farms are selected to represent three different climatic regions in the UK (Table 1). The climate is classified as temperate with cool winters and warm summers with wet weather throughout the year. Minimum and maximum temperatures occur in January and June, respectively.

\renewcommand{\arraystretch}{1.0}
\begin{table}[htb]
\centering
\caption{Climate of the studied farms}
\begin{tabular}{ | c | c | c | c | }
 \hline
 Farm & Region & Rainfall & Temperature \\ 
 \hline
 A & Saffron Walden, Essex & 545 & [-2.3,5.4] \\ 
 B & Salisbury, Wiltshire & 659 & [1.2,8.4] \\  
 C & Hull, Yorkshire & 488 & [2.2,9.6] \\
 \hline
\end{tabular}
\end{table}

Two fields having larger areas from each farm were selected for our spatial analysis. Location using UK grid reference, above see-level elevation, soil texture and cropping area of the selected fields are shown in Table 2. These codes given to each field are used to refer a field throughout the article.

\setlength{\tabcolsep}{9pt}
\renewcommand{\arraystretch}{1.0}
\begin{table}[htb]
\centering
\caption{Locations of the studied fields}
\begin{tabular}{ | c | c | c | c | c |}
 \hline
 ID &  Area & Elev & Soil \\ 
 \hline
 A1 &  23.2 & 109 & Clay soil \\ 
 A2 &  24.5 & 103 & Clay soil\\  
 B1 &  58.8 & 160 & Silty loam \\ 
 B2 &  60.2 & 140 & Silt soil\\  
 C1 &  38.9 & 268 & Sandy loam\\ 
 C2 &  40.1 & 405 & Sandy soil\\  
 \hline
\end{tabular}
\end{table}

We have the field boundary coordinates for each field. Based on that we create a 10mx10m base grid with a certain number of cells. It is vital that this grid is defined based on the field boundary map but not on the yield maps, which might change from year to another. This field specific time-independent grid (with cell centres) is used for interpolation and stacking yield maps from different years, cloud-free satellite imagery etc. into a single point. We also exclude the data points from a 20m wide border closer to the boundary, which mostly shows low yield and noise in NDVI due to fencing/boundary trees, left out for machinery movements etc., since our objective is to identify yield changing tendencies but not the outliers. 

\subsection{Yield Data}
Since 1992 when AgLeader introduced grain yield mapping technologies into their combine-harvesters, yield maps are generated by many such machineries. It took several more years (until selective availability restriction in GPS systems was decommissioned in 2000 for higher resolution) to integrate more accurate differential GPS into the mapping systems. The number of consecutive years the maps were available for the farmers was limited in the past 20 years. Therefore, those maps have vastly been under-utilised by the agri-tech companies, agronomists and the growers. Lack of end-user computational power and integrated decision-support systems for agronomists and farmers has also hindered new innovations to use such a valuable resource. 

For the fields we have selected yield monitoring maps are available for several consecutive years. For farm A, 7 maps from 2013 to 2019, for farm B, 10 maps from 2008 to 2017, and for farm C, 7 maps from 2013 to 2019 (except 2014) are available. A summary of the availability of the yield maps for those fields is shown in Table \ref{tab:yieldyear}. Table \ref{tab:yieldyear} also shows which years Winter Wheat was grown on the fields. Farm A has grown Barley Winter, Rape Winter and Beans Dried Winter in rotations with the Winter Wheat. Farm B has grown Barley Spring and Rape Winter in rotations. 

\setlength{\tabcolsep}{9pt}
\renewcommand{\arraystretch}{1.0}
\begin{table}[htb]
\centering
\caption{The studied yield maps}
\label{tab:yieldyear}
\begin{tabular}{ | c | c | l | }
 \hline
 Field & Maps & WW Growing Years \\ 
 \hline
 A1 &  7 & 2013, 2015, 2016, 2019, 2020 \\ 
 A2 &  7 & 2013, 2015, 2016, 2019, 2020 \\  
 B1 & 10 & 2014, 2017, 2020 \\ 
 B2 & 10 & 2015, 2018 \\  
 C1 & 7 & 2018 \\ 
 C2 & 7 & 2018, 2019 \\  
 \hline
\end{tabular}
\end{table}

Yield maps usually contain a set of errors and not uniformly sampled across a field since it logs data points according to the movements of the combine. Therefore, cleaning the yield maps to identify spatially contiguous set of different yield regions is important. The accuracy of any pre-processing workflow, depends on any approach for accurately identifying yield regions. We follow a combined temporal and spatial data pre-processing pipeline for filtering, smoothing, removing outliers, interpolations etc. to generate cleaned yield maps. 

First, we take a raw yield map and remove any data points specified to be excluded, if the combine-harvester yield mapping software has identified and indicated them as data points with errors (e.g., harvester not moving or moving outside the field, moving with header up, sudden stop etc.). We take only the data points within a specified range (e.g. from 1.0 to 16 T/Ha for Winter Wheat). Then we detect and remove global outliers based on the yield z-score values. We consider a data-point outside +/-3 z-score as outliers. Following this stage, we order the data-points according to the timestamp. As shown in Fig 1 (a), normally yield at the edges of a field are relatively low, when the combine-harvester take turns. A Hampel filter is used with imputations to identify the time-series outliers within a moving window and impute with average yield of the window. We also apply a moving-average filter (with a defined window size, 6 in the shown map) to smooth out the data dips along the time sequence of the combine-harvester. This window size is recommended to be adjusted according to the length of the field. 

Then, we apply ordinary kriging with a fitted exponential variogram (i.e. with estimated nugget, sill and effective range values, 0.42, 1.02 and 180m respectively in the shown map) to smooth and interpolate the original yield points into our uniformly spaced 10mx10m base grid centre points.  Then a convolutional spatial filter with 19m radius is applied to smooth the yield values (taking the local averages) across the field.  

\begin{figure}
     \centering
     \begin{subfigure}[b]{0.22\textwidth}
         \centering
         \includegraphics[height=1.5in]{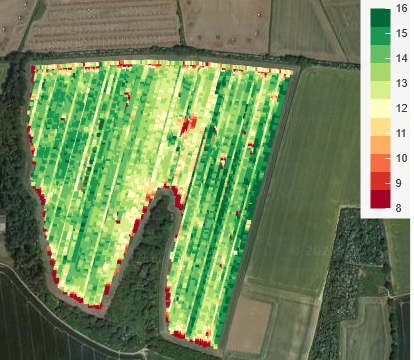}
         \caption{Raw Yield map}
         \label{fig: Raw Yield map}
     \end{subfigure}
     \hfill
     \begin{subfigure}[b]{0.22\textwidth}
         \centering
         \includegraphics[height=1.5in]{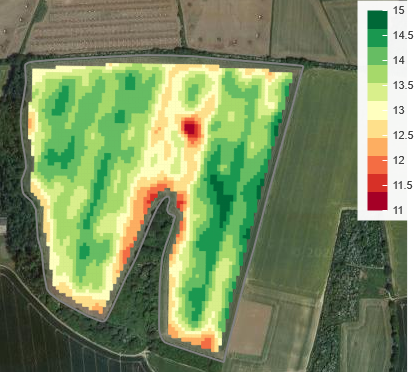}
         \caption{Pre-processed map}
         \label{fig: Pre-processed yield map}
     \end{subfigure}
     \hfill
        \caption{Combined temporal-spatial yield map pre-processing}
        \label{fig:Yield Map Pre-processing}
\end{figure}

Fig 1 (b) shows the improvement of spatial contiguity of yield maps after our combined temporal-spatial pre-processing workflow for the field B1. Mean yield has changed from 13.4 T/Ha to 13.6 T/Ha and standard deviation from 1.78 T/Ha to 0.63 T/Ha. It indicates that within field variations are smoothed to identify yield regions. Contiguity is improved as yield changes gradually from high to low across the regions. Map shown in Fig 1 (b) has 4532 10mx10m cells.

\subsection{Yield Frequency Map}
Contiguous historical yield potential and stability zones are delineated using a set of past yield maps. A yield frequency map is generated by stacking a set of yield maps. We use those frequency maps to understand persistent issues such as poor soil physical properties, soil compaction, waterlogging, poor seed establishment etc. within an area of a field. This indicate consistently low, high performing yield regions as well as unstable yield regions within a field (i.e. yield potential as well as yield stability). 
In order to generate a FM we normalise (min-max) yield data into the scale from -1 to +1. It supports crop rotation when we want to generate FM by stacking different crop types. Then we use Moran's LISA (Local Indicators of Spatial Association) clustering, which considers spatial dependency and heterogeneity into account. In the LISA scatter plot, High-Low, Low-High (as spatial outliers) and statistically not-significant regions are considered as unstable regions in a yield map and assigned 0. High-High regions having high yield with high correlation with neighbours are assigned +1 and Low-Low regions having low yield with high correlation with neighbours are assigned -1. Based on Moran yield we calculate yield frequency at a point \emph{i} according to the equation. 

\[ YF_{i} = \sum_{k=1}^{P} MC_{i} \]

The yield map of field B1 using all 10 years yield maps is shown in Fig 2 (a). It can range from +10 (consistently high) to -10 (consistently low). Fig 2 (b) shows the distribution of yield frequencies based on Moran LISA clustered (MC) yield. 
\begin{figure}
     \centering
     \begin{subfigure}[b]{0.18\textwidth}
         \centering
         \includegraphics[width=1.5in]{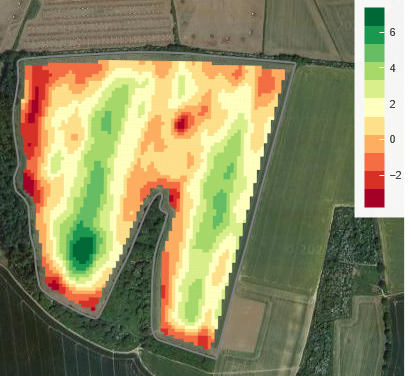}
         \caption{Frequency map}
         \label{fig: Raw Yield map}
     \end{subfigure}
     \hfill
     \begin{subfigure}[b]{0.28\textwidth}
         \centering
         \includegraphics[width=2.0in]{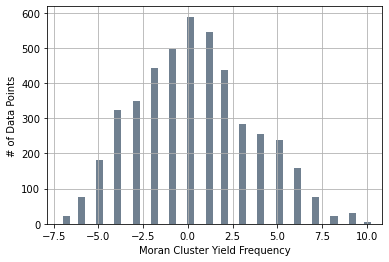}
         \caption{Frequency Distribution}
         \label{fig: Pre-processed yield map}
     \end{subfigure}
     \hfill
        \caption{Yield Frequency Map}
        \label{fig:Yield Map Pre-processing}
\end{figure}

\subsection{Satellite Imagery}
Remote sensing is one of most affordable way of using crop vigor monitoring in a large area. There are several commercial and non-commercial satellites now launched for agricultural purposes. MODIS, LandSat, Sentinel are non-commercial satellite services available at very low cost with good spatial resolutions. However, cloud cover is a major hurdle to use satellite images in the regions of northers Europe like UK. European Space Agency (ESA) Sentinel-2 satellite images are available since 2015 March from Sentinel-2A and since 2017 April from both Sentinel-2A and 2B. Therefore, in recent years its re-visit time is 2-3 days. 

\begin{figure}[!htb]
     \centering
     \begin{subfigure}[b]{0.22\textwidth}
         \centering
         \includegraphics[height=1.5in]{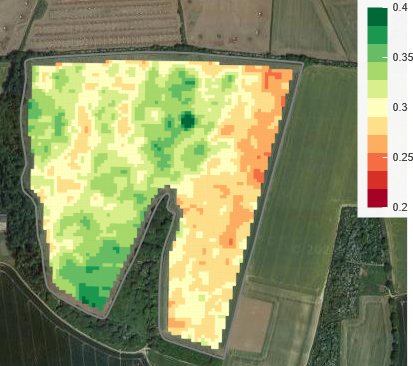}
         \caption{2020-01-20}
         \label{fig: Raw Yield map}
     \end{subfigure}
     \hfill
     \begin{subfigure}[b]{0.22\textwidth}
         \centering
         \includegraphics[height=1.5in]{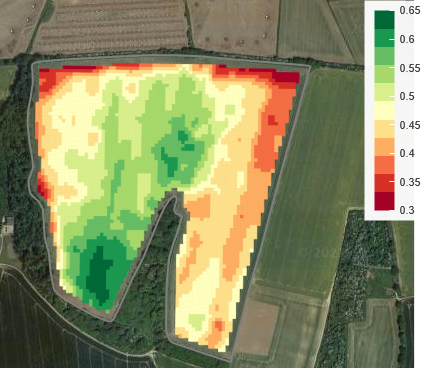}
         \caption{2020-03-25}
         \label{fig: Pre-processed yield map}
     \end{subfigure}
     \hfill
        \caption{Cloud-free NDVI images}
        \label{fig:Yield Map Pre-processing}
\end{figure}

We have acquired available images with field boundary as our area of interest with its could cover mask. Table 4 shows dates of the cloud free images for these fields within cropping years 2018, 2019 and 2020 (a cropping year, for example 2020, is considered from 01 September 2019 to 31 August 2020) when cloud cover probability is less than 10\% at a pixel and points within the field boundary after removing the 20m border.

Fig 4 (a) and (b) respectively shows the NDVI images of field B1 taken on cloud free days of 2020-01-20 and 2020-03-25. Winter wheat is at emerging (average NDVI = 0.309) and ripen (average NDVI = 0.483) stages respectively. Data pre-processing used the similar data workflow to the yield maps. However, we used Inverse Weighting Distance (IWD) for interpolation in the base grid.

\subsection{Vegetation Indexes}
There are several Vegetation Indexes (VIs) derived to describe crop vigor and hence the yield potential. Among them Normalised Difference Vegetation Index (NDVI), Enhanced Vegetation Index (EVI), Normalised Difference Red Edge (NDRE), Wide Dynamic Range Vegetation Index (WDRVI), Green Normalised Difference Vegetation Index (GNDVI) and Green Chlorophyll Vegetation Index (GCVI) are widely used with crop monitoring. Selecting highly correlated one with yield monitoring data is one of the evaluation we have considered in our paper.

Table 3 show how these indexes are derived using different bands in Sentinel-2. WDRVI is considered more accurate than NDVI when used with crop canopy monitoring. EVI is similar to NDVI but correct aerosol scattering as atmospheric condition and canopy background noise. This is more sensitive with dense canopy vegetation. GCI monitors leaf Chlorophyll using NIR and Green bands. GNDVI is a Chlorophyll based vegetation index but it saturates after NDVI. It detects water and nitrogen update of crop canopy. NDRE uses a narrow spectral range between visible red and NIR. It considered to be more accurate monitoring early and late stage of crop monitoring. Sentinel-2 provides Red Edge at 20m resolution while Red, Green, Blue, NIR are available at 10m spatial resolution. 

\setlength{\tabcolsep}{13pt}
\renewcommand{\arraystretch}{2.0}
\begin{table}
\centering
\caption{Vegetation Indexes of interest}
\begin{tabular}{ | c | c |  }
 \hline
 VI &  Equation  \\ 
 \hline
 NDVI & $ \frac{NIR-Red}{NIR+Red} $   \\ 
 EVI & $ \frac{2.5(NIR-Red)}{NIR+6Red-7.5Blue+1} $   \\  
 NDRE & $ \frac{NIR-RedEdge}{NIR+RedEdge} $   \\
 WDRVI & $ \frac{0.2NIR-Red}{0.2NIR+Red} $  \\
 GCVI & $ \frac{NIR}{Green} -1 $   \\
 GNDVI & $ \frac{NIR-Green}{NIR+Green} $   \\
 \hline
\end{tabular}
\end{table}

\begin{table*}[htb]
\centering
\caption{Satalite imagry availibility}
\begin{tabular}{|p{1cm}|p{4cm}|p{4cm}|p{4cm}|}
 \hline
 Field & 2018 & 2019 & 2020 \\ 
 \hline
 A1 & 04-19, 04-22, 05-07, 05-17, 05-19, 06-06, 07-01, 07-03, 08-02, 08-05, 08-07 & 03-05, 07-01, 08-25 & 09-14, 09-19, 09-21, 04-21, 04-26, 05-21, 05-28, 06-25, 07-20 \\ 
 A2 & 09-24, 10-06, 12-28, 05-17, 06-06, 06-18, 06-26, 08-02, 08-07 & 09-24, 09-26, 10-01, 05-14, 06-28 & 09-14, 09-19, 09-21, 03-22, 03-24, 04-26, 05-21, 05-28, 06-25 \\  
 B1 & 03-26, 04-20 & 02-26, 04-20, 06-29, 07-24 & 09-19, 05-29, 06-13, 07-30 \\ 
 B2 & 04-20, 05-07, 05-15, 06-29 & 07-24 & 09-17, 05-29, 08-12 \\  
 C1 & 04-13, 04-25, 05-17, 05-26, 06-21, 07-10, 08-02  & 03-10, 05-10, 06-25,07-10,08-15  & 09-09, 09-27, 10-15, 02-21, 03-26, 05-21, 06-28, 07-25, 08-20 \\ 
 C2 & 04-15, 04-28, 05-17,  07-10, 08-02  &  05-10, 06-25,07-10,08-15  &  10-15, 02-21, 03-26, 05-21, 06-28, 07-25, 08-20\\  
 \hline
\end{tabular}
\end{table*}

\lipsum[0]
\begin{figure*}
     \centering
     \begin{subfigure}[b]{0.28\textwidth}
         \centering
         \includegraphics[width=2.0in]{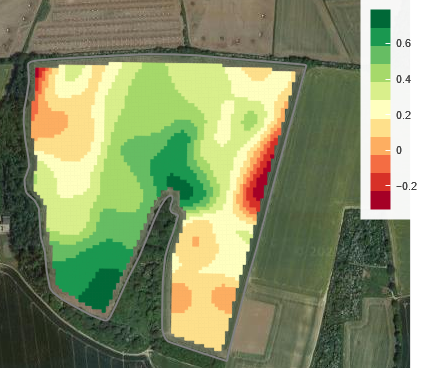}
         \caption{Intercept}
         \label{fig: Raw Yield map}
     \end{subfigure}
     \hfill
     \begin{subfigure}[b]{0.28\textwidth}
         \centering
         \includegraphics[width=2.0in]{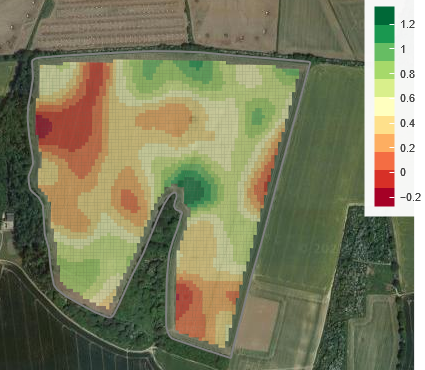}
         \caption{FM Coefficient}
         \label{fig: Pre-processed yield map}
     \end{subfigure}
     \hfill
     \begin{subfigure}[b]{0.28\textwidth}
         \centering
         \includegraphics[width=2.0in]{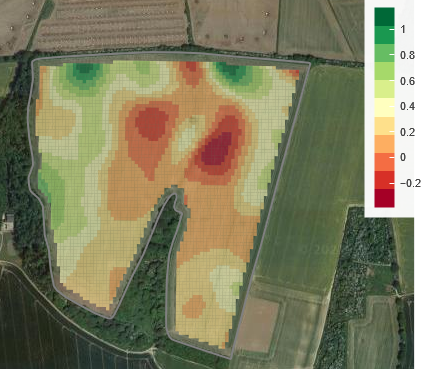}
         \caption{NDVI1 Coefficient}
         \label{fig: Pre-processed yield map}
     \end{subfigure}
     \hfill
        \caption{Spatially Varying Coefficients with NDVI and FM}
        \label{fig:Yield Map Pre-processing}
\end{figure*}

\subsection{Geographically Weighted Regression}
Geographically Weighted Regression (GWR) is a linear regression method applied locally by considering spatial auto-correlation. Regression algorithm is applied at a point with  a weight to other points. Weights of the points are determined by the proximity of the main point and use a Gaussian kernel function. As a result, points away from the main point are not or minimally considered. This process is applied all the points on the surface. As a result, we get a parameter surface instead of a single parameter. Therefore spatially varying regression equation determine that how important a feature is at this point.   

\[ y_i = \beta_i + \sum_{k=1}^{P} \beta_k x_{ik}+\epsilon_i \]

Where coefficient $\beta_i$ are the parameters. In global regression, these values are constant across the study area. However, if the variables are spatially auto-correlated, this equation hides this geographical richness of Tobler's phenomena. This spatial heterogeneity is needed to encounter in our model. As result Brunsdon et al. proposed GWR in 1996. We can use Moran I to assess spatial stationarity. GWR allows a model to vary in a contiguous way.

In this work, GWR is used with fixed bandwidth setting, gaussian-shaped spatial kernel and golden-search algorithm, which finds the optimal bandwidth. Bandwidth of 47m is detected in this field. Figure 3 shows the R2 accuracy surface of predicted NDVI map, coefficient surfaces of NDVI1 and YF. Predicted NDVI is highly accurate in most of the regions. If we consider area A in Figure 3 (b), it has a regression equation of NDVI2 = -0.8 + 1.3 NDVI1 + 0.1 YF. According to the equation NDVI2 is mainly driven by NDVI1 but not due to historical YF. This means it is a change detected to be addressed (incidental). In the area B as shown in Figure 3 (c), the regression equation is NDVI2 = 0.8 + 0.8 NDVI1 + 0.8 YF. In this case the contribution of YF is comparative and area is driven by historical yield performance (persistent). According to this approach we will delineate regions with persistent issues or incidental issues based on localised regression parameters. When subsequent NDVI images are available (March, April etc.) we re-apply GWR with the same process and refine the areas further for identifying changes.

\section{Software architecture and capabilities}

\begin{figure*}[!htb]
     \centering
\includegraphics[width=\textwidth]{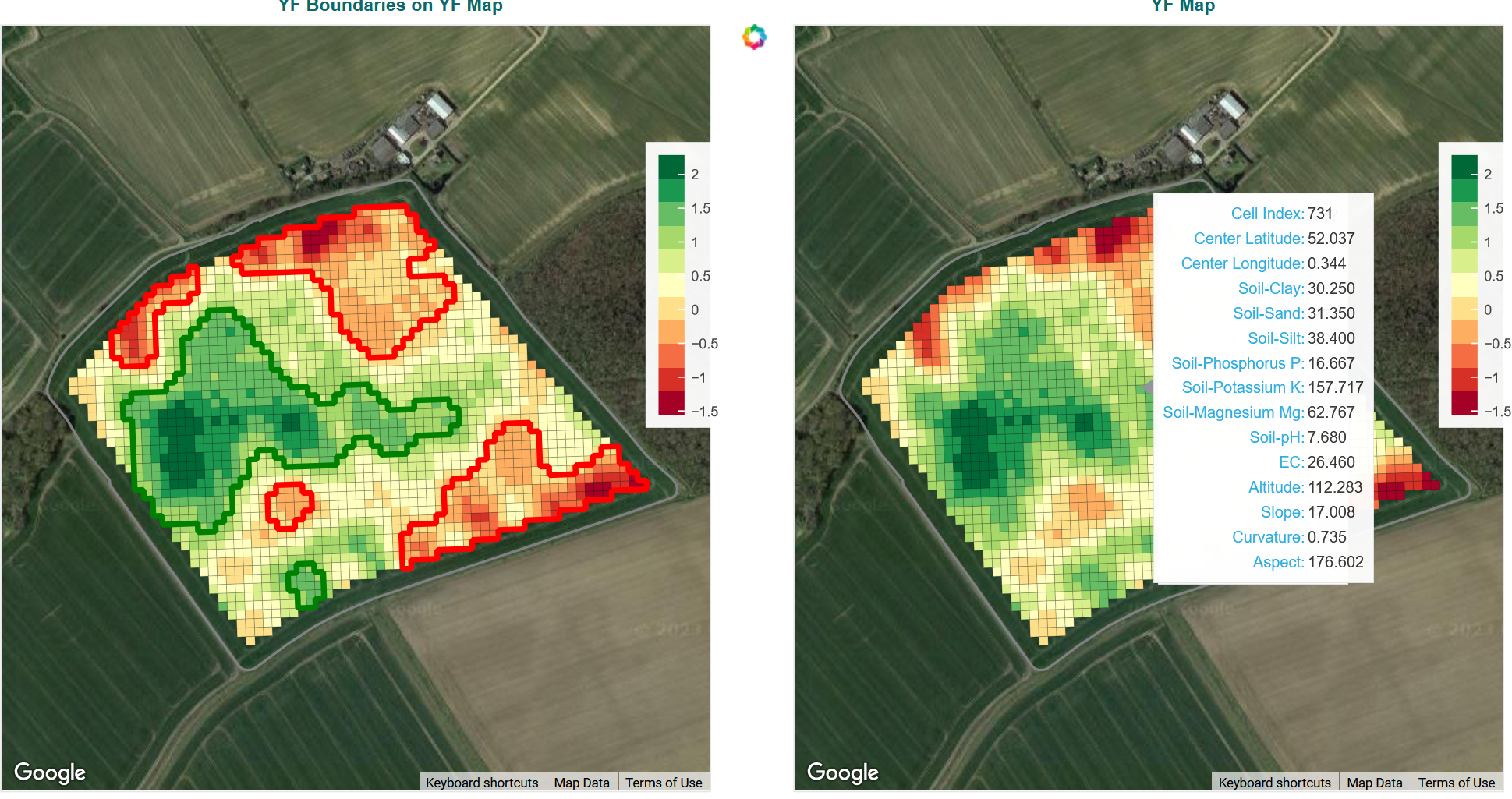}
\caption{Frequency map and zone monitoring}
\label{fig:yield_clustered}
\end{figure*}

In this section, we demonstrate the capabilities of the developed software, its software architecture and the main functions.

\subsection{Yield clustering and frequency maps}
As mentioned early, we apply clustering algorithms to analyse the yield distribution of a given field. Crop yield clustering plays a pivotal role in modern agriculture by enabling the early detection of abnormalities within fields. Through the analysis of data collected from various sensors and sources, ML algorithms can identify patterns and group similar crop yield data points together, forming clusters that represent different zones within the field. This technique empowers farmers to quickly pinpoint areas exhibiting unexpected variations in yield, which could indicate factors such as nutrient deficiencies, pests, diseases, or irrigation irregularities. By proactively identifying these anomalies, farmers can swiftly address and mitigate issues, optimizing resource allocation, reducing crop losses, and ultimately ensuring a more sustainable and productive agricultural operation. Figure \ref{fig:yield_clustered} shows an example of a frequency map generated by clustering the yield of a few years, which helps to identify problematic areas that exhibit low yields over the years. As we can see in the right side of Figure \ref{fig:yield_clustered}, the displayed map is interactive and can show the soil nutrient and field topological information of any cell in the field grid, each cell in the grid represents 10m by 10m. For each cell, the following information can be displayed: Cell Index, Center Latitude, Center Longitude, Soil-Clay, Soil-Silt, Soil-Sand, Soil nutrient (P, K, Mg ...etc.), Soil-Ph, Electric conductivity, Slope, Altitude, Curvature, Aspect.

\begin{figure}[!htb]
     \centering
\includegraphics[width=\columnwidth]{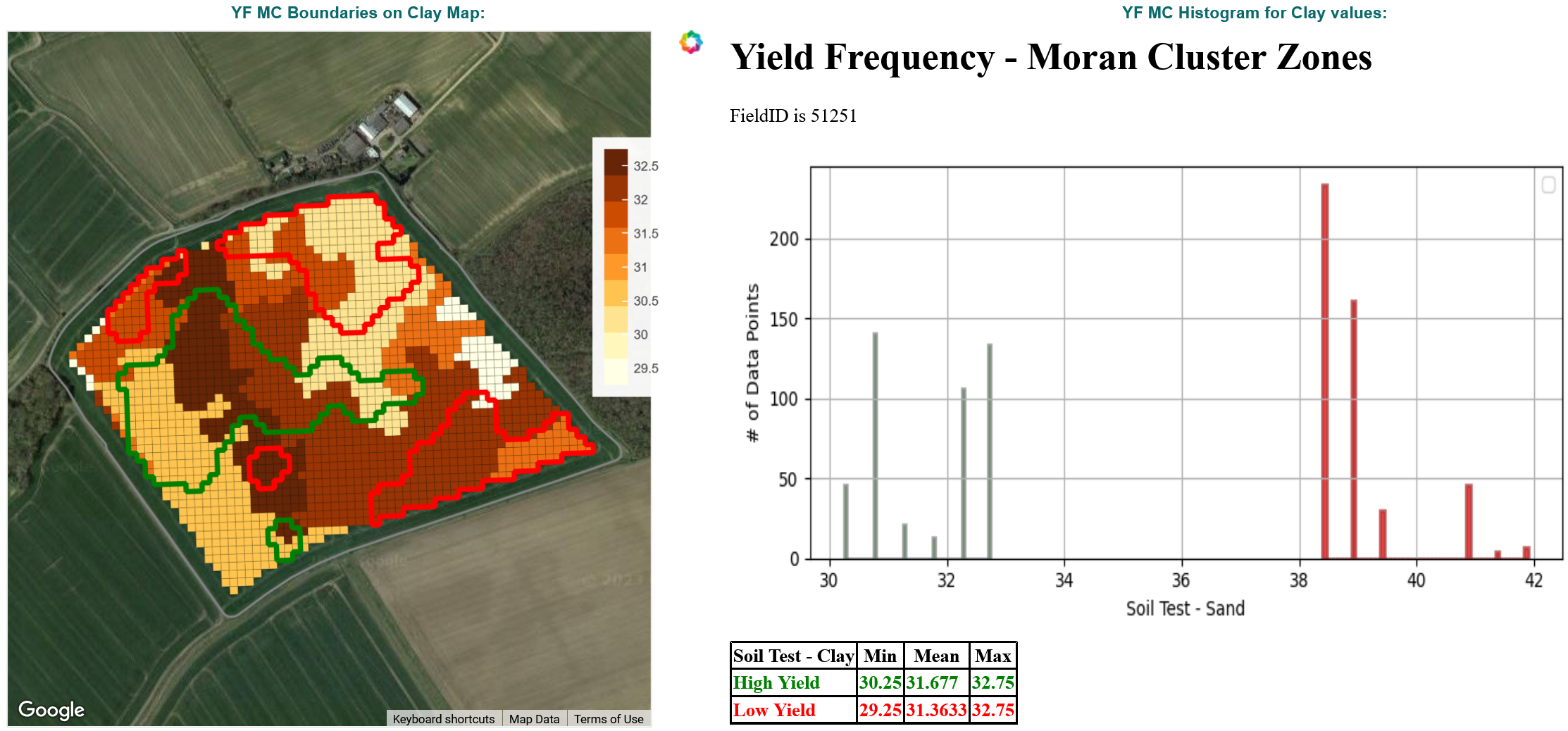}
\caption{Yield Frequency analysis using clay region distribution}
\label{fig:yf-clay}
\end{figure}

\subsection{Soil properties analysis}
Analyzing soil texture in terms of its composition of clay, sand, and silt particles holds immense significance for agriculture. The proportions of these components greatly influence the soil's water-holding capacity, drainage, nutrient retention, and aeration, directly impacting crop growth and yield. Clay-rich soils tend to retain water but can become compacted, affecting root penetration. Sandy soils drain quickly but may lack nutrients. Silt contributes to nutrient retention and soil structure. By understanding these relationships, farmers can tailor their cultivation practices, irrigation methods, and fertilizer applications to match the specific soil type. This knowledge allows for precise adjustments that optimize crop growth, minimize water usage, prevent nutrient loss, and ultimately enhance overall yield and quality, contributing to sustainable and efficient agricultural practices.

\begin{figure}[!htb]
     \centering
\includegraphics[width=\columnwidth]{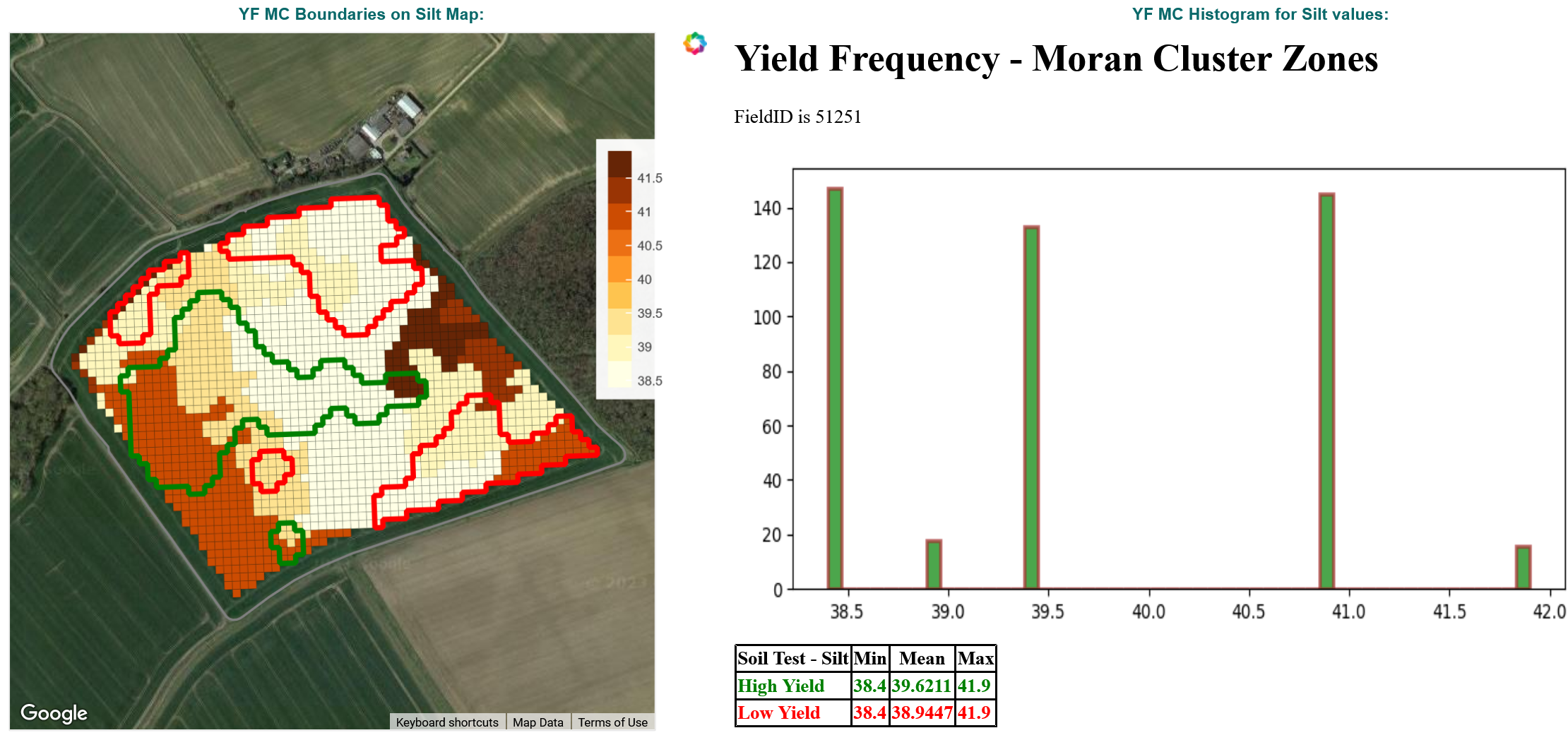}
\caption{Yield Frequency analysis using silt region distribution}
\label{fig:fy-silt}
\end{figure}

\begin{figure}[!htb]
     \centering
\includegraphics[width=\columnwidth]{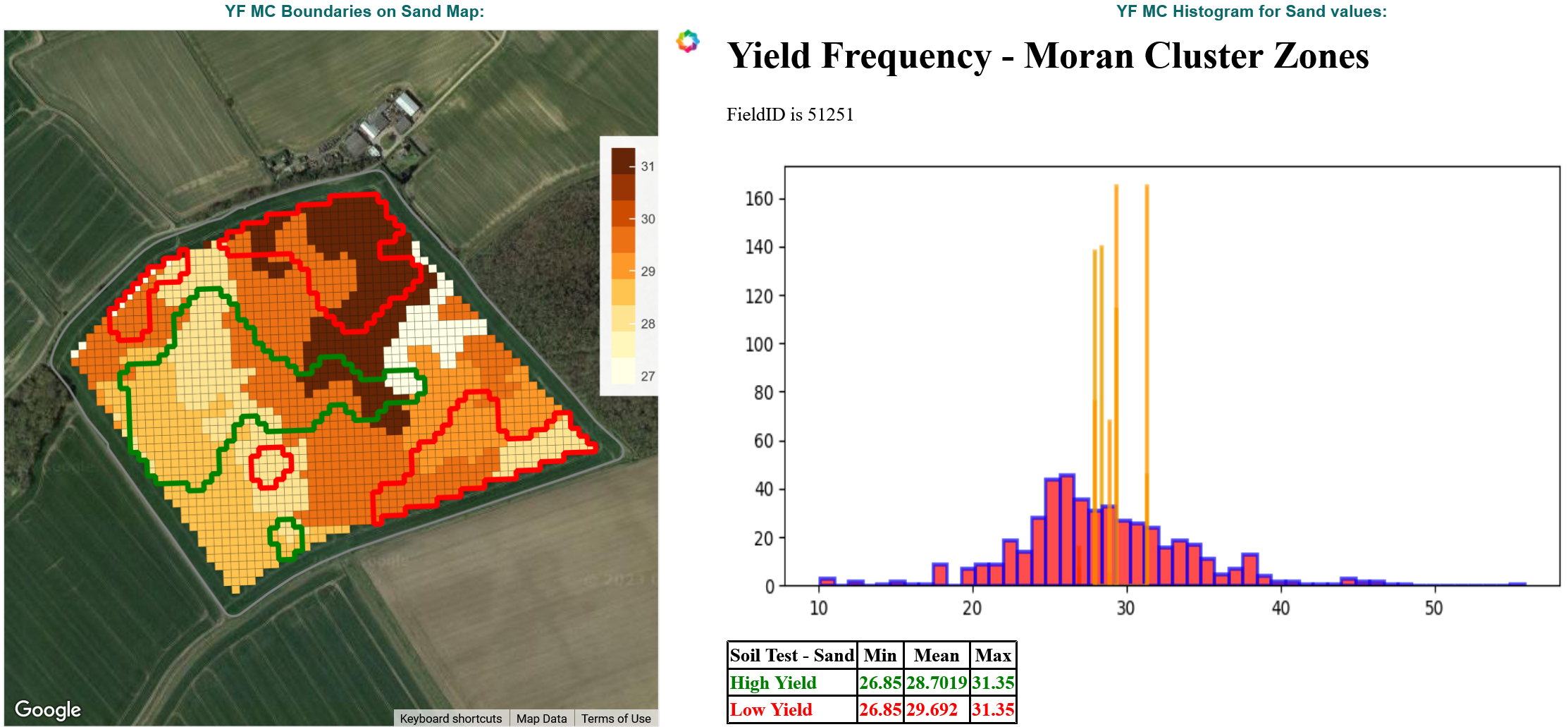}
\caption{Yield Frequency analysis using sand region distribution}
\label{fig:yf-sand}
\end{figure}

\begin{figure}[!htb]
     \centering
\includegraphics[width=\columnwidth]{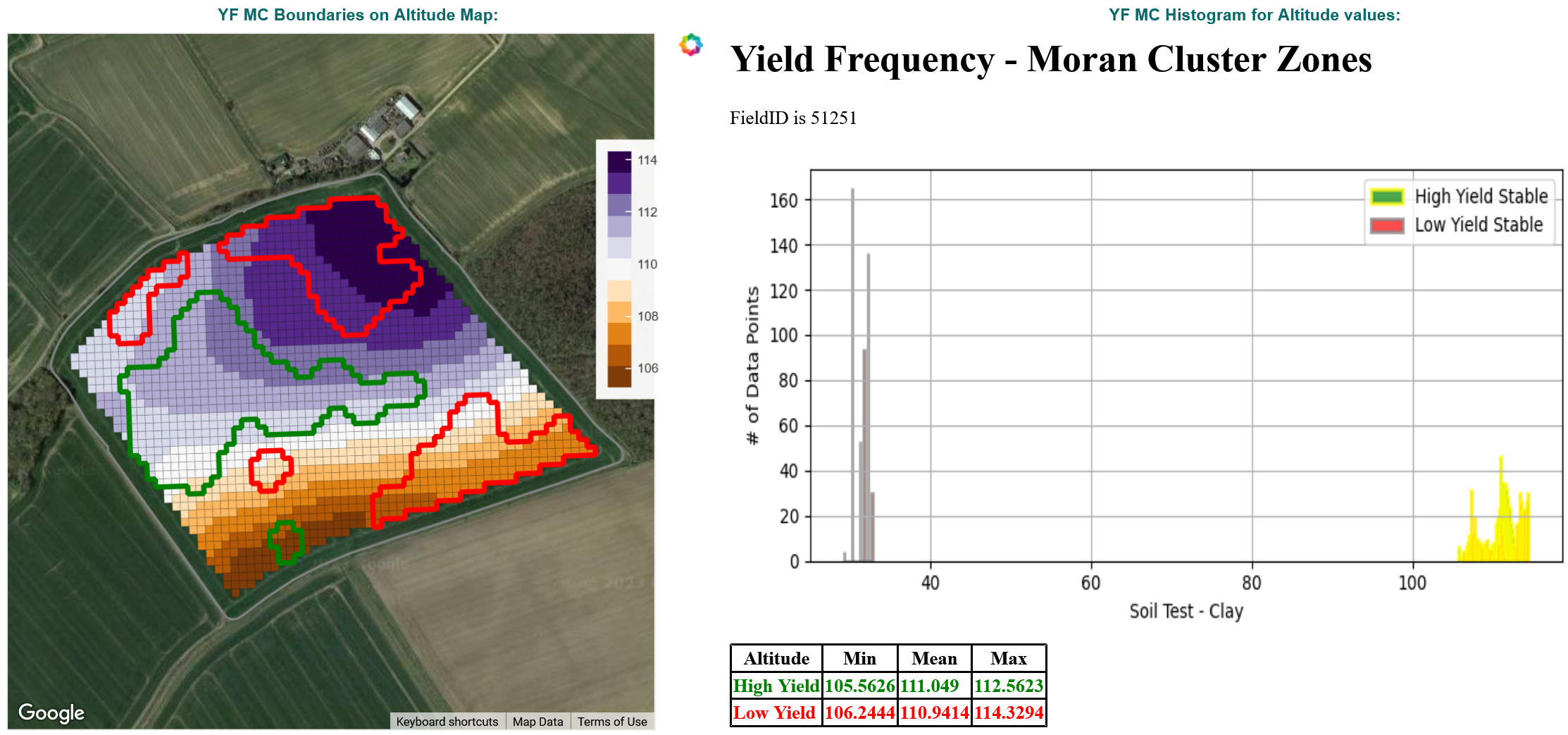}
\caption{Yield Frequency analysis in related to altitude}
\label{fig:yf-alt}
\end{figure}

\subsection{Spatial and topographic properties}
Analyzing the field's altitude, curvature, and exposure holds substantial importance in modern agriculture. These geographical factors impact the microclimate and overall environmental conditions within the field. Altitude influences temperature and sunlight exposure, while curvature and exposure affect water runoff and wind patterns. Understanding how these features interact with crop growth allows farmers to strategically plan crop placement and management. Higher elevations might offer cooler conditions ideal for certain crops, while south-facing slopes receive more sunlight. Additionally, lower areas could be prone to waterlogging. By factoring in these aspects, farmers can optimize planting times, choose suitable crop varieties, and adopt irrigation methods that align with the field's unique topography. This holistic approach maximizes yield potential, minimizes risk from adverse weather, and contributes to sustainable land use practices.
\begin{figure}[!htb]
     \centering
\includegraphics[width=\columnwidth]{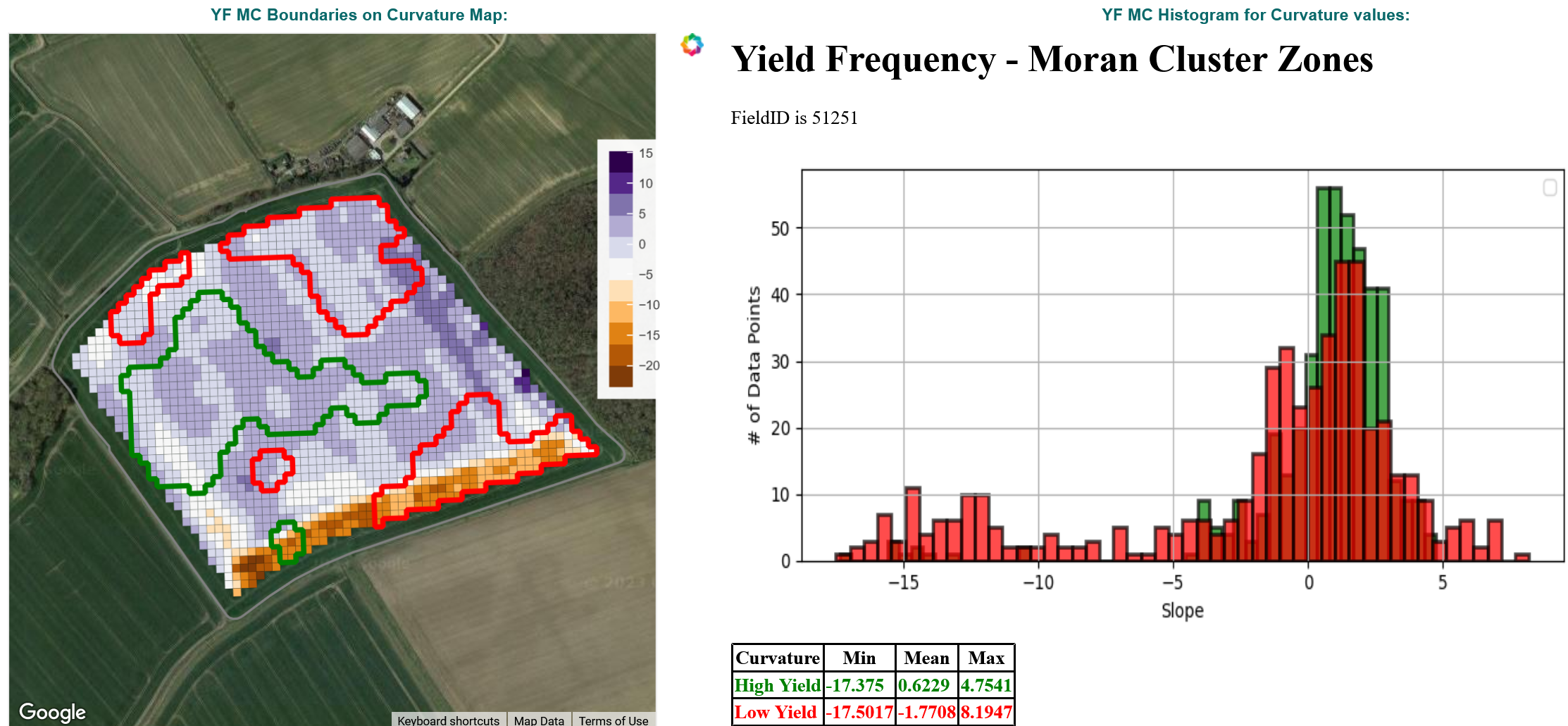}
\caption{Yield Frequency analysis in relation to curvature}
\label{fig:fy-curvature}
\end{figure}

\begin{figure}[!htb]
     \centering
\includegraphics[width=\columnwidth]{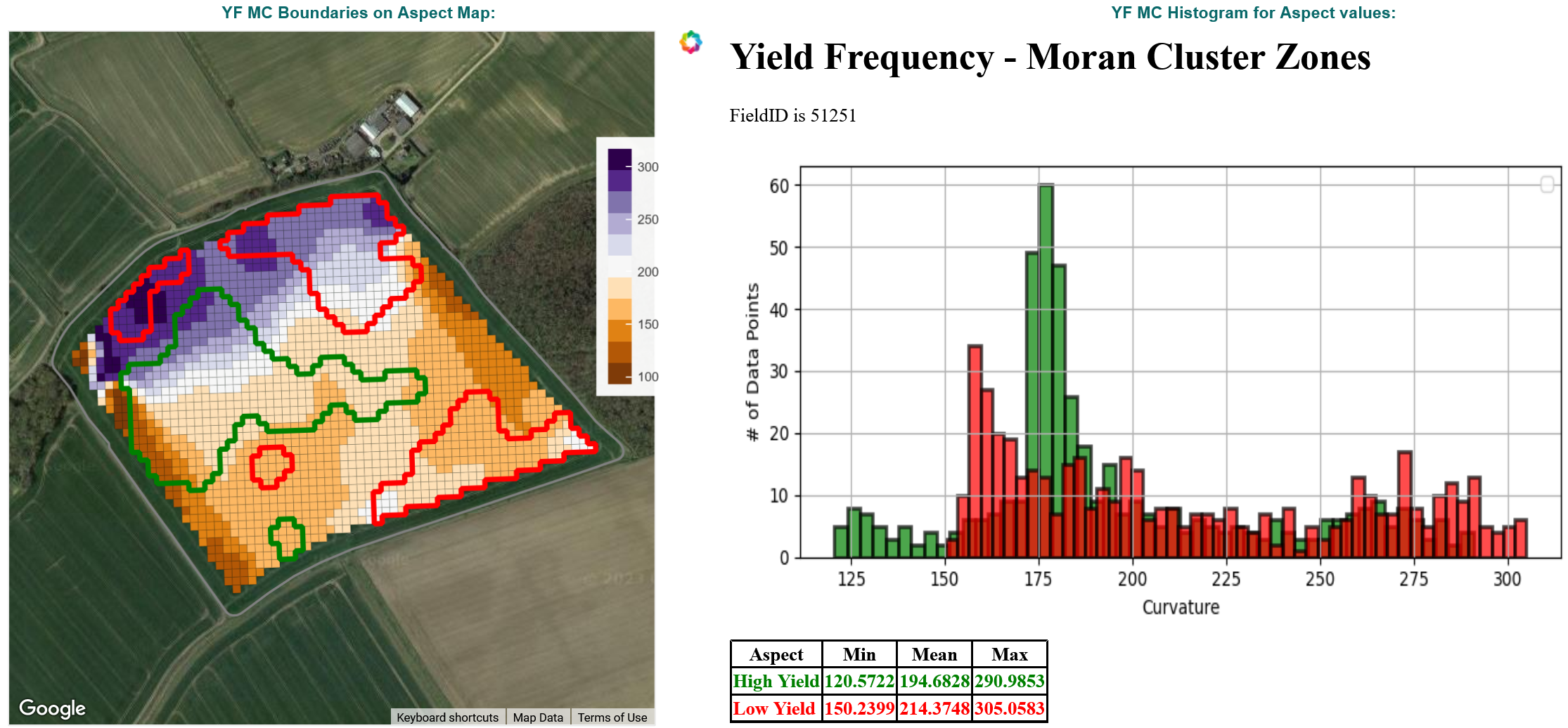}
\caption{Yield Frequency analysis in relation to aspect (exposure)}
\label{fig:yf-aspect}
\end{figure}

\section{Conclusion}

In this paper, we proposed a dynamic management zone delineation approach based on Geographically Weighted Regression using crop yield data, elevation and soil texture maps and available NDVI data. Our proposed dynamic management zone delineation approach is useful to analyse the spatial variation of yield zones. Delineation of yield regions based of historical yield data augmented with topography and soil physical properties helps farmers to economically and sustainably deploy site-specific management practices identifying persistent issues in a field. The use of frequency maps is capable of capturing dynamically changing incidental issues within a growing season. The proposed zone management approach can help farmers/agronomists to apply variable-rate N fertilisation more effectively by analysing yield potential and stability zones with satellite based NDVI monitoring.



\section*{Acknowledgment}

This is funded under the SFI Strategic Partnerships Programme (16/SPP/3296).





%
\bibliographystyle{elsarticle-num}
\bibliography{refs}


%

\end{document}